\RequirePackage{lineno}
\documentclass[preprint,superscriptaddress,aps,prl,superscriptaddress,showpacs,preprintnumbers,amsmath,amssymb]{revtex4}
\usepackage{graphicx}
\usepackage{xspace}
\usepackage{epsfig}
\usepackage{epstopdf}
\usepackage{multirow}
\usepackage{dcolumn}  % Align table columns on decimal point
\usepackage{wasysym}
\usepackage{subfigure}
\usepackage{color}
\usepackage{xr}
\usepackage{hyperref}
\usepackage{cleveref}
\usepackage{soul}
\usepackage{orcidlink}

\begin{document}

\preprint{\vbox{ \hbox{   }
							   \hbox{Belle Preprint 2022-18}
							   \hbox{KEK Preprint 2022-23}
%                        \hbox{Intended for {\it PRD RC}}
%                        \hbox{Author: Y.-T. Lai}
%                        \hbox{Committee: V. Savinov(chair),}
%                        \hbox{Y. Sakai, L. Piilonen, }
  		              % \hbox{hep-ex nnnn}
}}

%\linenumbers

\title{ \quad\\[1.0cm]\Large \bf \boldmath First Measurement of the $B^{+}\to\pi^{+}\pi^{0}\pi^{0}$ Branching Fraction and \textit{CP} Asymmetry}
\noaffiliation
  \author{Y.-T.~Lai\,\orcidlink{0000-0001-9553-3421}} % 2066
  \author{I.~Adachi\,\orcidlink{0000-0003-2287-0173}} % 2590
% \author{K.~Adamczyk\,\orcidlink{0000-0001-6208-0876}} % 2239
% \author{J.~K.~Ahn\,\orcidlink{0000-0002-5795-2243}} % 7423
  \author{H.~Aihara\,\orcidlink{0000-0002-1907-5964}} % 2223
  \author{S.~Al~Said\,\orcidlink{0000-0002-4895-3869}} % 6823
  \author{D.~M.~Asner\,\orcidlink{0000-0002-1586-5790}} % 4684
  \author{H.~Atmacan\,\orcidlink{0000-0003-2435-501X}} % 2538
  \author{V.~Aulchenko\,\orcidlink{0000-0002-5394-4406}} % 8183
  \author{T.~Aushev\,\orcidlink{0000-0002-6347-7055}} % 3747
  \author{R.~Ayad\,\orcidlink{0000-0003-3466-9290}} % 3766
% \author{T.~Aziz\,\orcidlink{-}} % 3523
  \author{V.~Babu\,\orcidlink{0000-0003-0419-6912}} % 5623
% ...
% 2387
% \author{V.~Zhulanov\,\orcidlink{0000-0002-0306-9199}} % 4983
\collaboration{The Belle Collaboration}

\noaffiliation

\begin{abstract}
%\begin{linenumbers}
We study $B^{+}\to \pi^{+}\pi^{0}\pi^{0}$ using 711 $\rm{fb}^{-1}$ of data collected at the $\Upsilon(4S)$ resonance with the Belle detector at the KEKB asymmetric-energy $e^{+}e^{-}$ collider. We measure an inclusive branching fraction of $(19.0\pm 1.5\pm 1.4)\times 10^{-6}$ and an inclusive \textit{CP} asymmetry of $(9.2 \pm 6.8 \pm 0.7)\%$, where the first uncertainties are statistical and the second are systematic; and a $B^{+}\to \rho(770)^{+}\pi^{0}$ branching fraction of $(11.2\pm 1.1\pm 0.9 ^{+0.8}_{-1.6})\times 10^{-6}$, where the third uncertainty is due to possible interference with $B^{+}\to \rho(1450)^{+}\pi^{0}$. We present the first observation of a structure around 1 GeV/$c^{2}$ in the $\pi^{0}\pi^{0}$ mass spectrum, with a significance of 6.4$\sigma$, and measure a branching fraction to be $(6.9\pm 0.9\pm 0.6)\times 10^{-6}$. We also report a measurement of local \textit{CP} asymmetry in this structure.

\pacs{14.40.Nd,13.25.Hw,13.25.-k,11.30.Er}
%\end{linenumbers}
\end{abstract}

\maketitle

Charmless three-body $B$ decays provide a rich environment to study the properties of the weak interaction in the quark sector~\cite{belle2phy}. 
The dynamics of such decays allows us to search for intermediate resonances and to study local \textit{CP} asymmetries~\cite{cp_dalitz_1}. These are important for developing better models to describe multibody hadronic $B$ decays. For $B\to 3\pi$, extraction of information on specific subdecay modes is useful for constraining phases of the Cabbibo-Kobayashi-Maskawa (CKM) matrix~\cite{ckm_1,ckm_2} elements.
For instance, the results for $B^{+}\to(\rho\pi)^{+}$ and time-dependent studies of $B^{0}\to(\rho\pi)^{0}$~\cite{charge_conjugate} allow determination of the CKM angle $\phi_{2}$~\cite{pentagon}. Also, interference between $B^{+}\to\chi_{c0}\pi^{+}$ and the nonresonant decays provides useful information for extracting the angle $\phi_{3}$~\cite{chic0}.

The first study of the inclusive branching fraction for $B^{+}\to\pi^{+}\pi^{0}\pi^{0}$ decays reported an upper limit of $8.9\times 10^{-4}$ at 90\% confidence level (C.L.)~\cite{pipi0pi0_argus}.
The branching fraction of $B^{+}\to\rho(770)^{+}\pi^{0}$ was measured by Belle~\cite{rhopi_1}, \textit{BABAR}~\cite{rhopi_2}, CLEO~\cite{rhopi_3}, and ARGUS~\cite{pipi0pi0_argus}. \textit{BABAR} and LHCb also performed amplitude analyses of $B^{+}\to\pi^{+}\pi^{-}\pi^{+}$ decays~\cite{pipipi_babar,pipipi_LHCb}, where intermediate resonances were investigated in detail.

In this Letter, we report measurements of the branching fraction and \textit{CP} asymmetry $\mathcal{A}_{CP}$ for $B^{+}\to \pi^{+}\pi^{0}\pi^{0}$. We use the $_splot$ technique~\cite{splot} to analyze the background-subtracted spectra, present the observation of a structure that is likely to be multiresonant, and measure its local \textit{CP} asymmetry. 
A major challenge in this Letter is the reconstruction of signal with two $\pi^{0}$ mesons, where the significant low-momentum (soft) $\pi^{0}$ background adversely affects our background-subtraction method.
%in which the low-momentum (soft) $\pi^{0}$ background has significant contribution adversely affecting our background subtraction method.

We use $772\times10^{6}$ $B\overline{B}$ pairs~\cite{NBB} collected at the $\Upsilon(4S)$ resonance with the Belle detector~\cite{Belle} at the KEKB asymmetric-energy $e^{+} e^{-}$ collider~\cite{KEKB}.

We use Monte Carlo (MC) samples to optimize selection criteria and determine the detection efficiency.
Samples of MC events for $\Upsilon(4S)\to B\overline{B}$ and hadronic continuum production $e^{+}e^{-} \to q\overline{q}~(q = u,d,s,c)$ are generated with {\mbox{\textsc{EvtGen}}\xspace}~\cite{ref:EvtGen} and simulated with {\mbox{\textsc{GEANT3}}\xspace}~\cite{geant}. 
For signal processes, we generate many MC samples for all relevant resonant decays, and nonresonant $B\to 3\pi$ decay distributed uniformly in the phase space.
All resonances are modeled by relativistic Breit-Wigner distributions.

Charged particles are reconstructed with the tracking detectors~\cite{Belle}. 
Reconstructed tracks' shortest distances to the interaction point (IP) are required to be within 5.0 cm along the $z$ axis (opposite the $e^{+}$ beam's direction), and within 0.3 cm in the transverse plane.  
We use information from particle identification detectors~\cite{Belle,PIDdetector} to calculate likelihood values $\mathcal{L}_{K}$ and $\mathcal{L}_{\pi}$ for kaon and pion hypotheses, respectively, for each track.
Tracks with $\mathcal{L}_{\pi}/(\mathcal{L}_{K}+\mathcal{L}_{\pi}) > 0.6$ are identified as pions. The efficiency for identifying a pion is 90\%; the probability to misidentify kaons as pions is less than 10\%.

The $\pi^{0}$ candidates are reconstructed from pairs of energy clusters, without associated track and reconstructed as photons, in the electromagnetic calorimeter~\cite{Belle}.
%To suppress beam-induced background, the reconstructed energy of each photon is required to be greater than 50 or 100 MeV if reconstructed in the barrel or end-cap regions, respectively. 
Beam-induced background is suppressed by requiring a photon energy above 50 or 100 MeV in the barrel or end cap regions, respectively. 
The invariant mass of each photon pair is required to be between 115 and 152 MeV/$c^{2}$, which is $\pm 3$ units of resolution around the known $\pi^{0}$ mass~\cite{PDG}.
To improve reconstruction of parent particles, kinematic fits, characterized by $\chi^{2}_{\pi^{0}}$, are performed for the $\pi^{0}$ candidates, constraining their invariant masses to the known $\pi^{0}$ mass~\cite{PDG}.

We form each $B^{+}$ candidate using a $\pi^{+}$ candidate and two $\pi^{0}$ candidates with distinct photons.
About 30\% of data events have more than one $B^{+}$ candidate, with an average candidate multiplicity of 1.6, primarily due to the soft $\pi^{0}$ background. 
We select the single $B^{+}$ candidate(s) in an event whose $\pi^{0}$ candidates have the smallest sum of their $\chi^{2}_{\pi^{0}}$ values; if multiple candidates remain, we select the $B^{+}$ candidate whose $\pi^{+}$ track has the shortest transverse-plane distance from the IP. 
In multicandidate events, this method selects the correct combination 92\% of the time, according to simulation.

To suppress the dominant background from hadronic continuum production, we use a neural network (NN)~\cite{NN} with inputs: a Fisher discriminant~\cite{Fisher} from 17 modified Fox-Wolfram moments~\cite{KSFW}; the cosine of the polar angle of the reconstructed $B$ direction and the cosine of the angle between the trust axis~\cite{thrust} of the reconstructed $B$ and that of the rest of the event, both in the c.m. frame; and the $B$ meson flavor tagging quality~\cite{flavortag}. 
The NN is trained with signal and continuum MC samples. Its output $C_{\mathrm{NN}}$ ranges from $-1$ to 1 and is required to be greater than 0.75. 
This retains 60\% of signal and removes 98\% of continuum background.
To simplify signal modeling, $C_{\mathrm{NN}}$ is transformed to $C^{'}_{\mathrm{NN}}\equiv \mathrm{log}(\frac{C_{\mathrm{NN}}-C^{\mathrm{min}}_{\mathrm{NN}}}{C^{\mathrm{max}}_{\mathrm{NN}}-C_{\mathrm{NN}}})$, where $C^{\mathrm{min}}_{\mathrm{NN}}$ is 0.75 and $C^{\mathrm{max}}_{\mathrm{NN}}$ is the maximum value of $C_{\mathrm{NN}}$ (obtained from the MC samples).

Background events from $B$ decays with the same final-state particles, $B^{+}\to \overline{D}^{0}\pi^{+}$ ($\overline{D}^{0}\to \pi^{0}\pi^{0}$) and $B^{+}\to K^{0}_{S} \pi^{+}$ ($K^{0}_{S}\to \pi^{0}\pi^{0}$), are removed by rejecting candidates with $M_{\pi^{0}\pi^{0}}$ within $\pm 3$ units of the $D^{0}$ or $K^{0}_{S}$ mass resolution around their known masses~\cite{PDG}.

Along with the correctly reconstructed (true) signal $B$ events in the signal MC samples, there is a sizable self-cross-feed (SCF) component arising from decay products of the other $B$ meson, primarily due to wrong photons or $\pi^{0}$'s included in signal reconstruction. 
Soft $\pi^{0}$ candidates in background events give rise to a structure in phase space that complicates the $_{s}plot$-based analysis since it distorts the $_{s}weights$ mass distributions.
To alleviate this problem, we require $p_{\pi^{0}}>$ 0.5 GeV/$c$ in the laboratory frame. %(as measured in the laboratory frame). 
This requirement reduces the efficiency of $B^{+} \to \rho(770)^{+}\pi^{0}$ by 35\%, while suppressing SCF by a factor of 2.

%%%%%%%%%%%%%%%%% 3D fit
We obtain the total signal yield and charge asymmetry $\mathcal{A}_{\mathrm{raw}}$ from a three-dimensional (the beam-energy constrained mass $M_{\rm bc}$, the energy difference $\Delta E$, and $C^{'}_{\mathrm{NN}}$) extended unbinned maximum-likelihood fit to data. $M_{\rm bc}$ is defined as $\sqrt{E^{2}_{\mathrm{beam}}/c^{4}-|\vec{p}_{B}/c|^{2}}$ and $\Delta E$ is defined as $E_{B}-E_{\mathrm{beam}}$, where $E_{\mathrm{beam}}$ is the beam energy 
and $\vec{p}_B$ and $E_B$ are the momentum and energy of the reconstructed $B^{+}$ candidate in the c.m. frame. The signal resolution is 3 MeV/$c^{2}$ for $M_{\mathrm{bc}}$ and 44 MeV for $\Delta E$. 
The likelihood function is
\begin{equation}
\mathcal{L}=\frac{e^{-\sum_{j}N_{j}}}{N!}\prod^{N}_{i=1}\left(\sum_{j}N_{j}P^{i}_{j}\right),
\end{equation}
where
\begin{equation}
P^{i}_{j}=\frac{1}{2}(1-q^{i}\mathcal{A}_{\mathrm{raw},j})\times P_{j}(M_{\rm bc}^{i},\Delta E^{i},C^{'i}_{\mathrm{NN}}).
\end{equation}
Here, $N$ is the number of candidate events, fit parameter $N_{j}$ is the expected number of  
events in category $j$, $q^{i}$ is the charge of the $\pi^{+}$ in the $i$th event, $\mathcal{A}_{\mathrm{raw},j}$ is the value of the charge asymmetry of the $j$th category, $P_{j}$ is the 3D probability density function (PDF) for category $j$, and $M_{\rm bc}^{i}$, $\Delta E^{i}$, and $C^{'i}_{\mathrm{NN}}$ are the values of these variables for the $i$th event. 
The fit region is $M_{\rm bc} >$ 5.26 GeV/$c^{2}$, $-0.3< \Delta E <$ 0.15 GeV, and $|C^{'}_{\mathrm{NN}}|<8$. 
We model the data with four event categories: signal, continuum, $B$ decays mediated via the dominant $b \to c$ transitions (``generic''), and $B$ decays mediated via $b \to u, d, s$ (``rare'').

Owing to shower leakage in the calorimeter~\cite{ecl}, $\Delta E$ and $M_{\rm bc}$ are correlated for signal events. 
Hence, the signal PDF is a 2D smoothed histogram in $\Delta E$ vs $M_{\rm bc}$ (obtained from MC events) multiplied by the sum of two Gaussian functions and an asymmetric Gaussian function representing $C^{'}_{\mathrm{NN}}$. The signal PDF includes both true signal and SCF contributions.
To correct for potential data-MC differences, signal PDF shapes are calibrated using a control sample of $B^{-}\to D^{0}\rho^{-}$, $D^{0}\to K^{-}\pi^{+}\pi^{0}$ decays.

The continuum background PDF is the product of an ARGUS function~\cite{argus} in $M_{\mathrm{bc}}$, a first-order polynomial in $\Delta E$, and the sum of two asymmetric Gaussian functions in $C^{'}_{\mathrm{NN}}$. 
Generic $B$ decays show no peaking structure in $M_{\mathrm{bc}}$ and $\Delta E$ after the $D^{0}$ veto, while rare $B$ decays, such as $B^{+} \to h^{+} \pi^{0}$ and $B^{0} \to \rho^{+} h^{-}$ ($h = \pi, K$), peak broadly in $M_{\mathrm{bc}}$ and have structure in $\Delta E$. 
To account for correlations, each $B\overline{B}$ background component is modeled using a 2D smoothed histogram in $\Delta E$ vs $M_{\rm bc}$ (obtained from simulation) multiplied by the sum of two Gaussian functions representing $C^{'}_{\mathrm{NN}}$. 
Except for the $\Delta E$ and $C^{'}_{\mathrm{NN}}$ shapes for continuum, the rest of the PDF shapes are fixed from simulation studies. To enhance the stability of the fitter, the parameters $\mathcal{A}_{\mathrm{raw}}$ are fixed to zero for backgrounds, which is consistent with MC predictions.

%%%%% iteration
We use the $_{s}weights$ obtained from the 3D fit to build the signal-isolated $M^{\rm min}_{\pi^{+}\pi^{0}}$ vs $M_{\pi^{0}\pi^{0}}$ histogram, where $M^{\rm min}_{\pi^{+}\pi^{0}}$ is the smaller of the two $M_{\pi^{+}\pi^{0}}$ values for a reconstructed $B^{+}$ candidate. 
In the $M_{\pi\pi}$ calculation, the momenta of the three pions are adjusted to constrain their total mass to the $B$ mass. We model the decays as an incoherent sum of subdecay modes and extract their yields from an extended weighted binned likelihood fit~\cite{roofit} to the 2D histogram, where the PDF of each subdecay is a 2D smoothed histogram taken from MC simulation of this subdecay.

We perform validation of the $_{s}plot$ approach as follows. After all the selection criteria are applied, the correlations between ($M_{\rm bc}$, $\Delta E$, $C^{'}_{\mathrm{NN}}$) and the three $M_{\pi\pi}$ are observed to be negligible. The $_{s}weights$ Dalitz plot distributions are confirmed to be consistent with signal yields in full simulation samples. 
Possible bias due to $_{s}\mathcal{P}lot$ in the 2D fit yields and statistical errors is studied using large toy MC ensembles and the data result. Variation in the results is taken as systematic uncertainty, with the details described later.

The signal PDF shape, yield of the 3D fit, the $_{s}\mathcal{W}eights$ distributions, and the 2D fit results are highly sensitive to the SCF fraction $F_{\mathrm{SCF}}$.
To simulate $F_{\mathrm{SCF}}$ correctly, we use an iterative procedure in which we generate new signal MC with a model based on the 2D fit result. 
We perform the 3D fit again with a new signal PDF obtained from the new simulated sample.
We perform five such iterations. The variation of $F_{\mathrm{SCF}}$ is less than 0.1\% between the last two iterations.

\begin{figure}[htb]
\centering
\includegraphics[width=0.42\textwidth]{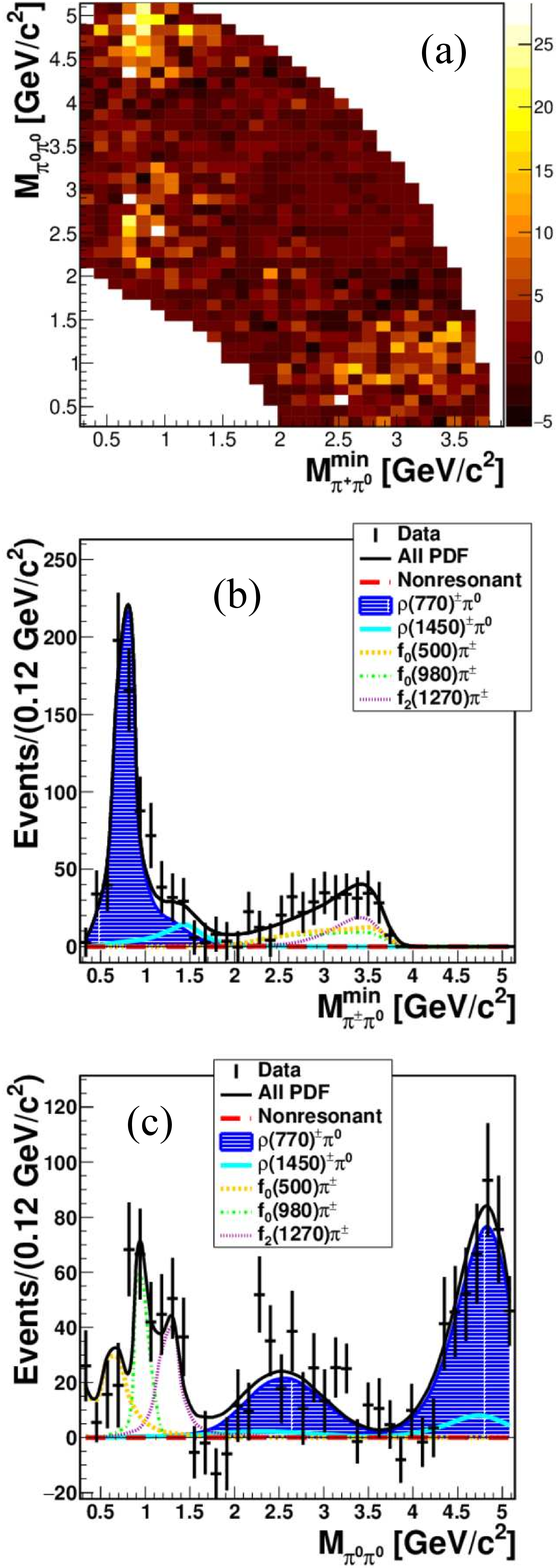}
\caption{$_{s}weights$ $M^{\rm min}_{\pi^{+}\pi^{0}}$ vs $M_{\pi^{0}\pi^{0}}$ distribution in (a), its projections, and the results of the 2D fit in (b) and (c).}
\label{fg:m2D} 
\end{figure}

%%%%%%%%%%%%%%%%% 2D fit result
From the final 3D fit, we obtain a signal yield of $1063\pm 86$ events and a raw asymmetry of $\mathcal{A}_{\mathrm{raw}}$ $(-9.2\pm 6.8)\%$ (see Supplemental Material~\cite{supp}), where the uncertainties are statistical.
Figure~\ref{fg:m2D}(a) shows the $_{s}weights$ $M^{\rm min}_{\pi^{+}\pi^{0}}$ vs $M_{\pi^{0}\pi^{0}}$ distribution with two broad clusters of events: one near the $\rho(770)^{+}$ resonance and the other around $M_{\pi^{0}\pi^{0}}=$ 1 GeV/$c^{2}$. The latter cannot be described by a single known resonance, so we model it by a sum of $f_{0}(500)$, $f_{0}(980)$, and $f_{2}(1270)$.

For the baseline model, we start with a sum of nonresonant $\pi^{+}\pi^{0}\pi^{0}$ and $\rho(770)^{+}\pi^{0}$, then include individual subdecays one by one in order of mass and repeat the 2D fit. Only modes that give a $p$ value of $F$ test~\cite{ftest} smaller than 0.5 are retained; $\chi^{2}$ is calculated with adaptive binning, requiring the number of entries of each bin to be greater than 1.5, where the number of degrees of freedom (d.o.f.) is the difference between the number of bins (127) and the number of PDFs in the 2D fit. The baseline model contains nonresonant $\pi^{+}\pi^{0}\pi^{0}$ decay, $\rho(770)^{+}\pi^{0}$, $f_{0}(500)\pi^{+}$, $f_{0}(980)\pi^{+}$, $f_{2}(1270)\pi^{+}$, and $\rho(1450)^{+}\pi^{0}$. 
Including each of $B^{+}\to f_{0}(1370)\pi^{+}$, $\chi_{c0}\pi^{+}$, or $\chi_{c2}\pi^{+}$ modes give a $p$ value of $F$ test greater than 0.5 and predicts a yield consistent with zero.
Figures~\ref{fg:m2D}(b) and (c) show $M^{\rm min}_{\pi^{+}\pi^{0}}$ and $M_{\pi^{0}\pi^{0}}$ projections of the data and fit results with the baseline model.
The model describes the data well with a $\chi^{2}$/d.o.f. of 0.93.

\begin{figure}[htb]
\centering
\includegraphics[width=0.42\textwidth]{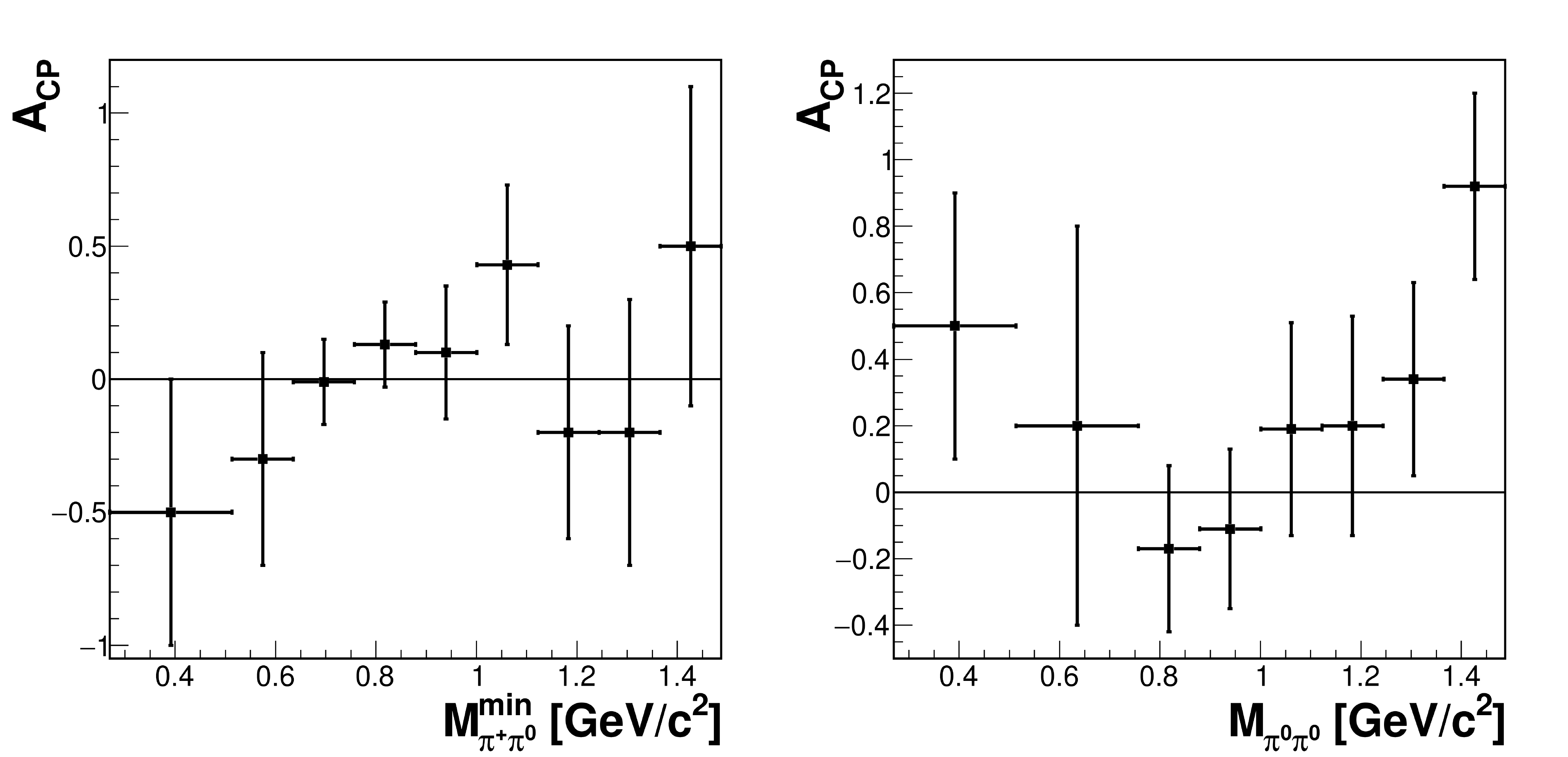}
\caption{$_{s}weights$ $\mathcal{A}_{CP}$ vs $M^{\rm min}_{\pi^{+}\pi^{0}}$ for $M_{\pi^{0}\pi^{0}}>$ 1.9 GeV/$c^{2}$, and vs $M_{\pi^{0}\pi^{0}}$ for $M^{\rm min}_{\pi^{+}\pi^{0}}>$ 1.9 GeV/$c^{2}$. The first few bins are combined due to small statistics.}
\label{fg:acp_mpi0pi0} 
\end{figure}

%%%%%%%%%%%%%%%%% BF
The branching fractions of $B^{+}\to \pi^{+}\pi^{0}\pi^{0}$ and its subdecays are 
\begin{equation}
\mathcal{B}=\frac{N_{\rm{sig}}}{\epsilon \times N_{B\overline{B}}},
\end{equation}
where $N_{\rm{sig}}$ and $\epsilon$ are the signal yield and reconstruction efficiency for each decay, and $N_{B\overline{B}}$ is the number of $B\overline{B}$ events. 
Here, true and SCF events are regarded as signal.
We assume equal branching fractions for the $\Upsilon(4S)$ decaying into charged and neutral $B\overline{B}$ pairs.
The efficiency for the inclusive decays is calculated using a MC sample of the final baseline model. $F_{\mathrm{SCF}}$ is 20.7\% for the inclusive decays and ranges from 10\% to 25\% for subdecay modes.

Table~\ref{tb:BF} summarizes our results. 
We do not report branching fractions for $f_{0}(500)\pi^{+}$, $f_{0}(980)\pi^{+}$, and $f_{2}(1270)\pi^{+}$ due to the significant overlap and insufficient interference information. As the region $M_{\pi^{0}\pi^{0}} <$ 1.9 and $M^{\rm min}_{\pi^{+}\pi^{0}}>$ 1.9 GeV/$c^{2}$ contains little contribution from nonresonant decay and $\rho$ modes, we report the branching fraction of $X\pi^{+}$, where $X$ stands for the total resonant $\pi^{0}\pi^{0}$ contribution, using the yield, $366^{+50}_{-48}$ events, of the 3D fit within that region. 
From the fit's likelihood distribution including all systematic uncertainties for $B^{+}\to X\pi^{+}$, $X\to\pi^{0}\pi^{0}$ mode as described below, the maximum likelihood $\mathcal{L}$ and the likelihood without the signal component, $\mathcal{L}_{0}$, are obtained. Using the value of $2~{\rm ln}(\mathcal{L}/\mathcal{L}_{0})$ and the change of two free parameters (yield and $\mathcal{A}_{\mathrm{raw}}$ of the signal), the corresponding significance of this yield is 6.4$\sigma$.

Upper limits at 90\% C.L. are reported for the modes with signals of statistical significance less than 3$\sigma$ using the frequentist method.
For each mode, we generate large toy MC ensembles from Table~\ref{tb:BF} and fit to these to obtain the yield distribution, where the overall systematic uncertainty for each mode is included by applying Gaussian smearing. We use these distributions to estimate the upper limits at 90\% C.L. Possible interferences between resonances are not included in this procedure. The asymmetry in $\pi^{+}$ detection is estimated using a control sample of $D^{+}\to K^{0}_{S}\pi^{+}$~\cite{kspi} as $-0.03\%$, and is subtracted from $\mathcal{A}_{\mathrm{raw}}$ to calculate $\mathcal{A}_{CP}$.

\begin{table*}[t]
\begin{center}
\caption{Summary of the masses and widths (in MeV/$c^{2}$) used in the 2D fit, signal selection efficiencies, fitted yields, branching fractions, and $\mathcal{A}_{CP}$. The values in parentheses are the upper limits of branching fraction at 90 CL. 
The first error is statistical and the second is systematic. The interference effect is included as the third uncertainty in the analysis of $\rho(770)^{+}\pi^{0}$ mode.}
\begin{tabular}{lcccccc}
\hline \hline
Decay mode & Mass & Width &$\epsilon$ (\%) & Fitted yield & $\mathcal{B}$ ($10^{-6}$) & $\mathcal{A}_{CP}$ (\%) \\
\hline
$\pi^{+}\pi^{0}\pi^{0}$ (inclusive) & & & 7.2 & $1063\pm 86$ & $19.0\pm 1.5\pm 1.4$ & $9.2\pm 6.8 \pm 0.7$ \\
\hline
Nonresonant &  &  & 10.6 &  $3\pm 14$ & $0.03\pm 0.16^{+0.12}_{-0.15}$ ($<0.6$) & $\cdots$  \\
$\rho(770)^{+}\pi^{0}$, $\rho(770)^{+}\to\pi^{+}\pi^{0}$ & 775.5 & 150.3 & 7.3 & $637 \pm 65$ & $11.2\pm 1.1\pm 0.9 ^{+0.8}_{-1.6}$ & $8.0 \pm 15.0 ^{+2.3}_{-7.5}$  \\
$\rho(1450)^{+}\pi^{0}$, $\rho(1450)^{+}\to\pi^{+}\pi^{0}$ & 1465 & 400 & 8.6 & $80\pm 51$ & $1.2\pm 0.6\pm 0.2$ ($<2.5$)& -- \\
$f_{0}(500)\pi^{+}$, $f_{0}(500)\to\pi^{0}\pi^{0}$ & 600 & 400 & 7.1 & $123\pm 37$ & $\cdots$ & $\cdots$ \\
$f_{0}(980)\pi^{+}$, $f_{0}(980)\to\pi^{0}\pi^{0}$ & 980 & 50 & 8.7 & $102\pm 30$ & $\cdots$ & $\cdots$ \\
$f_{2}(1270)\pi^{+}$, $f_{2}(1270)\to\pi^{0}\pi^{0}$ & 1275.4 & 185.1 & 5.6 & $119\pm 32$ & $\cdots$ & $\cdots$ \\
$X\pi^{+}$, $X\to\pi^{0}\pi^{0}$ & $\cdots$ & $\cdots$ & 6.9 & $366^{+50}_{-48}$ & $6.9\pm 0.9\pm 0.6$ & $18.2\pm 11.6\pm 0.7$ \\
$f_{0}(1370)\pi^{+}$, $f_{0}(1370)^{0}\to\pi^{0}\pi^{0}$ & 1400 & 300 & 9.0 & $<75$ & $<1.1$ & $\cdots$ \\
$\chi_{c0}\pi^{+}$, $\chi_{c0}\to\pi^{0}\pi^{0}$ & 3415.2 & 10.2 & 11.1 & $<39$ & $<0.5$ & $\cdots$ \\
$\chi_{c2}\pi^{+}$, $\chi_{c2}\to\pi^{0}\pi^{0}$ & 3556.3 & 2.0 & 11.5 & $<63$ & $<0.7$ & $\cdots$ \\
\hline \hline
\end{tabular}
\label{tb:BF}
\end{center}
\end{table*}

The local \textit{CP} asymmetries obtained from the $_{s}weights$ histograms for $B^{+}$ and $B^{-}$ are shown in Fig.~\ref{fg:acp_mpi0pi0}. The regions above 1.49 GeV/$c^{2}$ are not shown as the signal yields are consistent with zero. 
$\mathcal{A}_{CP}$ is consistent with 0 everywhere except for the $M_{\pi^{0}\pi^{0}}$ region between 1.36 and 1.49 GeV/$c^{2}$, which has $\mathcal{A}_{CP}=(92\pm 28)\%$. By performing an additional 3D fit to this region with $\mathcal{A}_{CP}$ floated in the range $[-1,1]$ and $\mathcal{A}_{CP}$ fixed to zero, we calculate a statistical significance of 3.2$\sigma$ for the nonzero local $\mathcal{A}_{CP}$ in this region. It is similar to the asymmetry in $B^{+}\to f_{2}(1270)\pi^{+}$ observed for $B^{+}\to\pi^{+}\pi^{-}\pi^{+}$~\cite{pipipi_babar,pipipi_LHCb}.

Various sources of systematic uncertainties are considered for all branching fractions. To obtain the overall value for each decay mode, all relevant independent uncertainties are summed quadratically. 
The reconstruction efficiency is calibrated for data-MC discrepancies using dedicated control samples; the small corrections are applied and their uncertainties are taken as systematic uncertainties.  
The uncertainty due to the number of $B\overline{B}$ events is 1.4\%. The uncertainty due to charged-track reconstruction is 0.35\% per track from $D^{*+}\to D^{0} \pi^{+}$ with $D^0 \to \pi^{+} \pi^{-} K^{0}_{S}$. The uncertainty due to $\pi^{+}$ identification is 0.9\% from $D^{*+}\to D^{0}\pi^{+}$ with $D^{0}\to K^{-}\pi^{+}$. The uncertainty due to $\pi^{0}$ reconstruction is 4.8\% from $\tau\to\pi^{-}\pi^{0}\nu_{\tau}$~\cite{pi0sys}. The uncertainty due to continuum suppression based on $C_{\mathrm{NN}}$ is 1.4\% from $B^{-}\to D^{0}\rho^{-}$ with $D^{0}\to K^{-}\pi^{+}\pi^{0}$. The uncertainty in estimating the reconstruction efficiency due to the MC statistics is 0.02\%.

To estimate the systematic uncertainty associated with fixed PDF shapes in the 3D fit, 
we vary the shapes of the signal PDF and analytic-function PDFs according to their respective uncertainties, and vary the binning schemes of all the other smoothed histograms. The resulting changes in the signal yield are added in quadrature.
To account for possible data-MC difference on $F_{\mathrm{SCF}}$ in the signal PDF, we vary it within $\pm 30\%$ of nominal.
The total systematic uncertainty is 5.5\%.

Possible variation of the signal model composition is estimated from the iteration procedure. We take the 0.7\% difference between yields in the last two iterations as a systematic uncertainty. Possible bias due to the $_{s}plot$ technique is estimated to be 2.9\% from the difference between the 3D fit yield and the sum of $_{s}weights$ in the limited Dalitz region.

While validating the entire fit procedure using large sample of pseudoexperiments, small biases are identified in both 3D and 2D fits. They result in (0.04--0.18)$\times 10^{-6}$ changes in branching fractions and are included as systematic uncertainties. 
The uncertainty due to each resonance's parameters is estimated. The mass and width are varied by $\pm 1$ unit of their uncertainties or over their entire range~\cite{PDG}; all changes in the yield are added in quadrature and taken as a systematic uncertainty. For the nonresonant decay PDF, the effect of varying its template's binning is studied. Since the nonresonant decay and $\rho(1450)^{+}\pi^{0}$ are the two least significant components in the 2D fit, we also consider the nonresonant decay's yield discrepancy with and without $\rho(1450)^{+}\pi^{0}$ PDF. The uncertainty range is between $0.003\times 10^{-6}$ and $0.42\times 10^{-6}$.

A systematic uncertainty in the efficiency for the inclusive decay due to decay-model uncertainties is estimated to be 1.6\% from the difference between the nominal branching fraction and its value obtained by summing over an efficiency-corrected $_{s}weights$ yields for all bins of $M^{\rm min}_{\pi^{+}\pi^{0}}$.
The uncertainty due to the $X\to\pi^{0}\pi^{0}$ model composition is estimated to be 2.9\% from the change in the efficiency when varying the fitted yields of $f_{0}(500)\pi^{+}$, $f_{0}(980)\pi^{+}$, and $f_{2}(1270)\pi^{+}$ by $\pm 1$ unit of their uncertainties.

The uncertainty in the $B\to X\pi^{+}$ branching fraction due to nonresonant and $\rho$ contamination is 3.3\% from the difference between the 3D fit yield and the sum of the 2D fit yields of the three resonances.

To estimate the systematic uncertainty due to interference between $\rho(770)^{+}$ and $\rho(1450)^{+}$, large toy MC ensembles with different phase differences and amplitude ratios are generated.
We fit to them using the incoherent sum of the PDFs for the two $\rho$'s and take the largest deviation between the fitted and the input amplitude ratios as a systematic uncertainty.

Several sources of systematic uncertainty for $\mathcal{A}_{CP}$ are considered.
The uncertainty due to $\pi^{+}$ detection asymmetry is 0.3\% using a control sample of $D^{+}\to K^{0}_{S}\pi^{+}$~\cite{kspi}. 
An uncertainty of 0.5\% due to fixing the PDF shapes in the 3D fit for the overall $\mathcal{A}_{CP}$ is studied using methods similar to that used for the branching fraction by varying the PDF shapes.
Similarly, the uncertainty due to resonance shape parameters in the 2D fit is $^{+2.2}_{-7.5}\%$ for $\rho(770)^{+}\pi^{0}$ by varying resonance shape parameters. 
The uncertainty due to the fixed background $\mathcal{A}_{CP}$ in the 3D fit is 0.5\% from the change in the results when its amount is floated. 
A systematic effect on $\mathcal{A}_{CP}$ due to interference between $\rho(770)^{+}\pi^{0}$ and $\rho(1450)^{+}\pi^{0}$ is negligible as determined from a study similar to that for the branching fraction.

In conclusion, we have performed a study of $B^{+}\to\pi^{+}\pi^{0}\pi^{0}$ using 711 fb$^{-1}$ of data collected by Belle. 
We measure the inclusive branching fraction $\mathcal{B}(B^{+}\to\pi^{+}\pi^{0}\pi^{0})=(19.0\pm 1.5 \pm 1.4)\times 10^{-6}$ and \textit{CP} asymmetry $\mathcal{A}_{CP} = (9.2 \pm 6.8 \pm 0.7)$\%, where the first uncertainties are statistical and the second are systematic.
We report the composition of intermediate states in $B^{+}\to\pi^{+}\pi^{0}\pi^{0}$ within our model and measure the local \textit{CP} asymmetry.
The branching fraction of $B^{+}\to\rho(770)^{+}\pi^{0}$ is measured to be $(11.2\pm 1.1\pm 0.9 ^{+0.8}_{-1.6})\times 10^{-6}$, where the third uncertainty accounts for possible interference with $B^{+}\to\rho(1450)^{+}\pi^{0}$. 
We observe a structure, likely arising due to multiple resonances, at $M_{\pi^{0}\pi^{0}}<$ 1.9 and $M^{\rm min}_{\pi^{+}\pi^{0}}>$ 1.9 GeV/$c^{2}$ with an inclusive branching fraction of $(6.9\pm 0.9\pm 0.6)\times 10^{-6}$. 
We report a measurement of local \textit{CP} asymmetry at 1.36 $<M_{\pi^{0}\pi^{0}}<$ 1.49 and $M^{\rm min}_{\pi^{+}\pi^{0}}>$ 1.9 GeV/$c^{2}$.
We do not observe $B^{+}\to f_{0}(1370)\pi^{+}$, $B^{+}\to \chi_{c0}\pi^{+}$, or $B^{+}\to\chi_{c2}\pi^{+}$. 
An amplitude analysis with improved treatment of systematic effects from $\pi^{0}$ reconstruction is needed to further understand the properties of the $B^{+}\to\pi^{+}\pi^{0}\pi^{0}$ transition, especially for the structure at low $M_{\pi^{0}\pi^{0}}$. 
Eventually, the larger dataset and better performance for neutral particle reconstruction at Belle~II~\cite{belle2,belle2phy} will enable such an analysis.

This work, based on data collected using the Belle detector, which was
operated until June 2010, was supported by 
the Ministry of Education, Culture, Sports, Science, and
Technology (MEXT) of Japan, the Japan Society for the 
Promotion of Science (JSPS), the Tau-Lepton Physics 
Research Center of Nagoya University, and the Kavli Institute for the
Physics and Mathematics of the Universe of University of Tokyo
established by World Premier International Research Center Initiative
(WPI); 
the Australian Research Council including Grants No.~DP180102629, % Sevior
No.~DP170102389, % Varvell
No.~DP170102204, % Yabsley
No.~DE220100462, % Hsu
No.~DP150103061, % Urquijo
No.~FT130100303; % Urquijo;
Austrian Federal Ministry of Education, Science and Research (FWF) and
FWF Austrian Science Fund No.~P~31361-N36;
the National Natural Science Foundation of China under Contracts
No.~11675166,  %Wen-Biao Yan
No.~11705209,  %Yi-Ming Li
No.~11975076,  %Chengping Shen
No.~12135005,  %Chengping Shen 
No.~12175041,  %Xiaolong Wang
No.~12161141008; %Chengping Shen
Key Research Program of Frontier Sciences, Chinese Academy of Sciences (CAS), Grant No.~QYZDJ-SSW-SLH011; % Chang-Zheng Yuan
Project ZR2022JQ02 supported by Shandong Provincial Natural Science Foundation;
the Ministry of Education, Youth and Sports of the Czech
Republic under Contract No.~LTT17020;
the Czech Science Foundation Grant No. 22-18469S;
Horizon 2020 ERC Advanced Grant No.~884719 and ERC Starting Grant No.~947006 ``InterLeptons'' (European Union);
the Carl Zeiss Foundation, the Deutsche Forschungsgemeinschaft, the
Excellence Cluster Universe, and the VolkswagenStiftung;
the Department of Atomic Energy (Project Identification No. RTI 4002) and the Department of Science and Technology of India; 
the Istituto Nazionale di Fisica Nucleare of Italy; 
National Research Foundation (NRF) of Korea Grant
No.~2016R1\-D1A1B\-02012900, No.~2018R1\-A2B\-3003643,
No.~2018R1\-A6A1A\-06024970, No.~RS\-2022\-00197659,
No.~2019R1\-I1A3A\-01058933, No.~2021R1\-A6A1A\-03043957,
No.~2021R1\-F1A\-1060423, No.~2021R1\-F1A\-1064008, No.~2022R1\-A2C\-1003993;
Radiation Science Research Institute, Foreign Large-size Research Facility Application Supporting project, the Global Science Experimental Data Hub Center of the Korea Institute of Science and Technology Information and KREONET/GLORIAD;
the Polish Ministry of Science and Higher Education and 
the National Science Center;
the Ministry of Science and Higher Education of the Russian Federation, Agreement 14.W03.31.0026, % from 15.02.2018
and the HSE University Basic Research Program, Moscow; % from 15.04.2021
University of Tabuk research Grants No.~S-1440-0321, No.~S-0256-1438, and No.~S-0280-1439 (Saudi Arabia);
the Slovenian Research Agency Grants No.~J1-9124 and No.~P1-0135;
Ikerbasque, Basque Foundation for Science, Spain;
the Swiss National Science Foundation; 
the Ministry of Education and the Ministry of Science and Technology of Taiwan;
and the U.S. Department of Energy and the National Science Foundation.
These acknowledgements are not to be interpreted as an endorsement of any
statement made by any of our institutes, funding agencies, governments, or
their representatives.
We thank the KEKB group for the excellent operation of the
accelerator; the KEK cryogenics group for the efficient
operation of the solenoid; and the KEK computer group and the Pacific Northwest National
Laboratory (PNNL) Environmental Molecular Sciences Laboratory (EMSL)
computing group for strong computing support; and the National
Institute of Informatics, and Science Information NETwork 6 (SINET6) for
valuable network support.


\begin{thebibliography}{99}
\bibitem{charge_conjugate} Throughout this Letter, inclusion of charge-conjugate decay modes is implied.
\end{thebibliography}
\end{document}